\documentclass{article}
\usepackage[a4paper,top=3cm,bottom=3.5cm,left=3.2cm,right=3.2cm]{geometry}

\usepackage[utf8]{inputenc}
\usepackage{authblk}
\usepackage{graphicx}
\usepackage{subfigure}
\usepackage{amssymb}
\usepackage{lineno}
\usepackage[english]{babel}
\usepackage{palatino, url}
\usepackage{amsmath}

\usepackage{listings}
\usepackage{graphicx}
\usepackage[rightcaption]{sidecap}
\usepackage{textcomp}
\usepackage{array} 
\usepackage{booktabs} 
\usepackage{multirow} 
\usepackage{caption} 
\usepackage{floatflt}
\usepackage{rotating} 
\usepackage{scrextend}
\usepackage{multirow}
\usepackage[numbers]{natbib}
\usepackage{xcolor}
\usepackage{hyperref}
\hypersetup{                                
    colorlinks=true,                
    breaklinks=true,                
    urlcolor= blue,                
    linkcolor= magenta,                              
    bookmarksopen=false,
    filecolor=black,
    citecolor=blue,
    linkbordercolor=blue
}

\title{Data informed epidemiological-behavioural modelling}

\date{\small To appear on the \textit{Proceedings of the GIMC-SIMAI Young 2024 Conference} (July 2024, Naples, IT), within the \textit{Lecture Notes in Mechanical Engineering} edited by Springer.}

\author[1]{Daniele Proverbio}
\author[1]{Riccardo Tessarin}
\author[1]{Giulia Giordano}
\affil[1]{\small Department of Industrial Engineering, University of Trento, Trento 38123, IT }

\begin{document}

\maketitle

\begin{abstract}
Augmenting classical epidemiological models with information from the social sciences helps unveil the interplay between contagion dynamics and social responses. However, multidisciplinary integration of social analysis and epidemiological modelling is often challenging, due to scarcity of vast and reliable data sources and because ad hoc modelling assumptions may not reproduce empirically observed patters.
Here, we test the hypothesis that awareness and information spreading straightforwardly translates into behavioural responses, analysing empirical data to generate insights about their dynamics and relationships. We employ such results to build a data-informed behavioural-epidemiological model that elucidates the impact of compliant behaviours and the role of centralised regulations in mitigating epidemics. We investigate model properties and its benefits in integrating theoretical modelling and data. 
\end{abstract}

\section{Introduction}
Mathematical modelling of infectious diseases is crucial to elucidate epidemic dynamics \cite{Brauer2017,Giordano2020}, enable predictions \cite{Kemp2021,Proverbio2021} and develop control strategies \cite{Alamo2021,Vargas2022}. 
Disease transmission is affected not only by infection dynamics, but also by socio-economical aspects \cite{Burzyński2021}, as well as by
behavioural responses to disease and information spreading \cite{Bauch2013}, which play a key role for epidemic spreading and interventions \cite{Gosak2021,Tyson2020}. To capture this effect, models from the social sciences need to inform epidemic modelling \cite{Bedson2021,Usher2020}.

A first approach to integrate social and epidemiological aspects employs mean-field homogeneous compartment models, such as the Susceptible-Infectious-Removed (SIR) model \cite{Kermak1927}, with time-varying parameters that depend on the spreading process \cite{Fenichel2011} and integrate behavioural aspects implicitly. 
An opposite approach involves agent-based or game-theoretic models \cite{Ye2021} that explicitly consider multiple aspects of disease transmission and population dynamics \cite{Palomo2022}. These, however, require large and heterogeneous datasets to be of practical effectiveness \cite{Proverbio2024}. Alternatively, augmented SIR-like models with behavioural compartments \cite{Bulai2023}, or models on networks \cite{Frieswijk2022} 
and multi-layer networks \cite{Peng2021}, have been recently developed to incorporate different aspects of epidemic-behavioural feedback. However, they are often based on ad-hoc assumptions and hardly relate to empirical data -- especially since behavioural data during epidemics are particularly challenging to collect and interpret as dynamical time series \cite{Nunner2021}. To overcome these limitations,
proxy models rely on compartmental or network models, which are calibrated on proxy data for behaviours, such as awareness \cite{Zino2021} 
or opinions \cite{Anderson2019}, 
which can be extracted e.g. from social networks and are likely to translate into behaviours -- but not necessarily, nor directly. Also, assuming linear dynamics for the entire behavioural response could be too restrictive to accurately represent the influence of behaviours during epidemics \cite{Huys2011}.

In this work, we aim at bridging  mathematical modelling and empirical data on the evolution of behaviours during an epidemic. To this aim, we analyse a recent dataset about worldwide behavioural responses, collected during the COVID-19 pandemic, and leverage upon it to build a dynamical epidemiological-behavioural model. We use data-driven reasoning to set the model time scales and parameter ranges, and we study the model properties and evolution, elucidating the impact of behavioural elements and paving the way to studying dynamical patterns in coupled epidemiological-behavioural dynamics.

\section{Data collection and analysis}

To shed light onto the use of proxy data and behavioural dynamics, inform the development of our model and identify realistic ranges for model parameters, we analyse a recent dataset, created during the COVID-19 pandemic \cite{Astley2021}. We analyse correlations, patterns and timescales between social and behavioural features. 

\subsection{Methods}

For social and behavioural data, we use the dataset developed by the University of Maryland Social Data Science Center ``Global COVID-19 Trends and Impact Survey" initiative, in partnership with Facebook. The dataset consists of users' answers to questionnaires, run on the Facebook social network in most countries from 2020 to 2022. It contains over 100k daily responses, which were anonymised, weighted, normalised, curated, aggregated and made available through API by the University of Maryland. 
The full description of data collection and curation methodologies can be found in \cite{Barkay2020}. As any web survey, the dataset presents some limitations and biases, discussed on the project website \\\texttt{gisumd.github.io/COVID-19-API-Documentation/}. Still, it is one of the most complete collections of time-series, and it has been abundantly used to uncover correlations and trends of behavioural responses during the pandemic \cite{Heino2023}. 

Data consist in an array of indicators, pertaining to specific domains -- symptoms, behaviour, economic, mental health, etc. Each indicator is associated with a counter, representing the (normalised) number of respondents who answered positively to its associated question. For instance, the indicator ``mask'' (behavioural indicator) counts how many respondents (in \%) wore a mask all the time or most of the time when in public. 
All indicators and their explanation are listed on the project website. 

The database is accessed via its Python API. Among the indicators, we select those related to awareness, trust on communicators, beliefs, and behaviours, on the topic of mask wearing. Our choice of indicators was corroborated by a trained behavioural scientist. We focus on mask wearing since it is one of the most covered topics in the database, and its debate was arguably less polarised than that on vaccines over the considered time frame, which spanned from 21$^{st}$ May 2021 to 25$^{th}$ June 2022 (for which data were consistently available for all indicators).

Since progresses from awareness to behaviours, particularly within fluid contexts with both peer pressure and guidelines from central institutions, may differ depending on cultures and geographies \cite{Astley2021}, a universal model cannot be developed, and tailoring to context is needed. Also, Facebook coverage is not homogeneous around the world, and the dataset curators also acknowledged heterogeneous response patters. So, we focus on three European countries: Italy, Germany and United Kingdom (mainland). They all display widespread adoption of the social network among all age groups, have an average respondents' rate of about 1\%, and are sufficiently similar with respect to cultural and behavioural aspects -- including reactions to laws and guidelines from authorities.

We collect data on the time evolution of the epidemic from the John Hopkins University repository \cite{dong2020interactive}, \texttt{https://github.com/CSSEGISandData/COVID-19}, as incidence data (daily confirmed cases) and confirmed deaths. Data regarding policies and their stringency are 
based on the Oxford COVID-19 Government response tracker and Stringency Index \cite{Hale2021}. These  data are collected from the beginning of the pandemic, in late February 2020, until July 2022.

\subsection{Results}

\begin{figure}[t]
	\centering
	\includegraphics[width=0.8\linewidth]{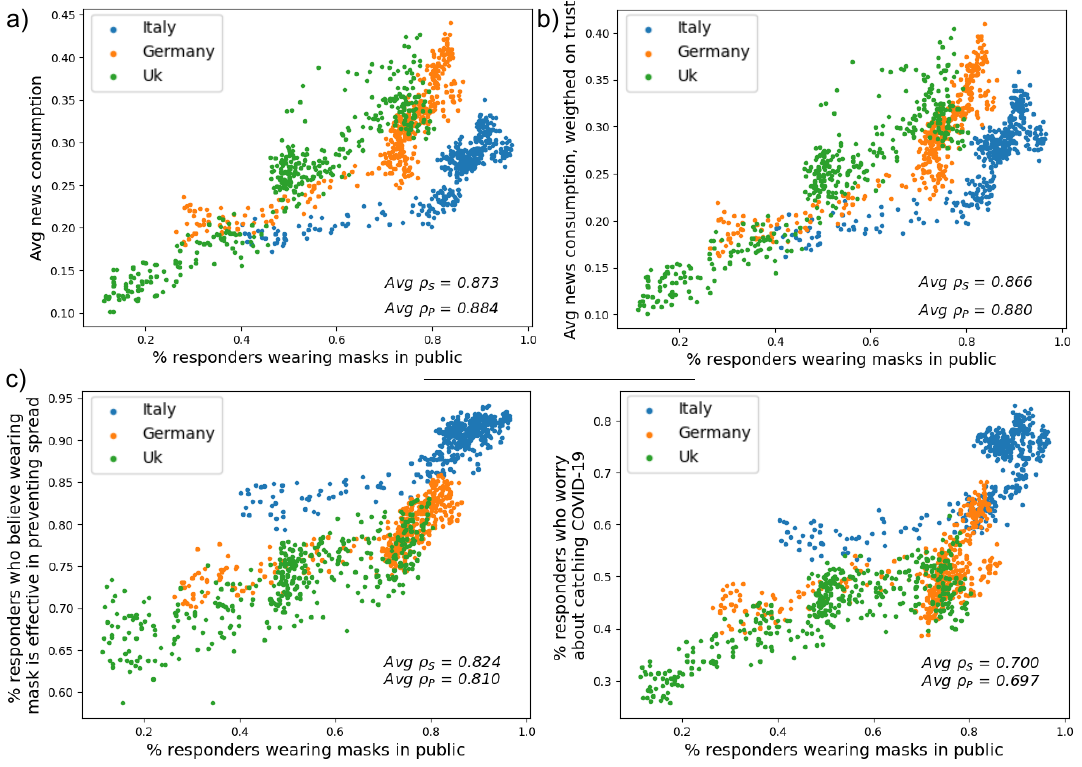}
	\caption{\small Correlation between mask wearing behaviour and: (a) awareness (mean exposure to news sources); (b) awareness weighted on trust; (c) two types of beliefs. Spearman's $\rho_S$ and Pearson's $\rho_P$ correlation coefficients are reported (bottom-right).}
	\label{fig:awa-beh}
\end{figure}

To begin with, we verify that behaviours do not perfectly correlate over time with awareness, as hypothesised by previous studies \cite{Nunner2021}. As a case study, we use as first variable $x[t]$ the indicator referring to the behaviour of ``wearing a mask all times or most of the times when in public''. The second variable $y_1[t]$, referring to average awareness about the epidemic status and about the measures enacted in a country, is given by $y_1 = N^{-1} \sum_{i=1}^N \theta_i[t]$, where $\theta_i$ are the indicators related to exposure to news sources (local health workers or clinics, scientists and health experts, govt. health authorities/officials, WHO, friends and families, journalists, politicians), regardless of the medium. The result is reported in Fig. \ref{fig:awa-beh}a: on average, mask-wearing behaviour correlates rather monotonously with awareness (Spearman's coefficient $\rho_S = 0.87$, p-value = $2 \cdot 10^{-100}$), but not perfectly in a linear fashion (Pearson's correlation $\rho_P = 0.88$). This also depends on country, with UK having $\rho_P$ as high as 0.94 and Italy as low as 0.82. These results suggests that, on average, using proxy data based on awareness may be an initially good approximation to calibrate models. However, it is best to directly employ behavioural data and related dynamics to overcome the limitations of inaccuracies stemming from nonlinear relationships between personal stances.

We also test the effect of trust, by computing the correlation between $x[t]$ and $y_2[t] = (\sum_i w_i[t])^{-1} \sum_{i=1}^N \theta_i[t] w_i[t]$, where $w_i$ is the percentage of respondents saying that they trust the $i$-th news source. This way, each news source is weighted by the trust placed upon it. The result, in Fig. \ref{fig:awa-beh}b, suggests that trust does not play a significant role in modifying the average behavioural response to news exposure. In fact, the correlation statistics are almost equal to the unweighted case, and similar considerations apply. 

Finally, we test the mediating role of beliefs. In fact, 
behaviours are mediated by beliefs, which process information from the environment (awareness) into action. To this end, we check whether $x[t]$ correlates with $z_1[t]$, i.e., the percentage of respondents believing that wearing masks is effective in preventing spread, and $z_2[t]$, related to being worried about catching COVID-19. The results, reported in Fig. \ref{fig:awa-beh}c, suggest that beliefs about efficacy correlate 
to the immediate awareness of pandemic unfolding, while being worried contributes to a lesser degree. 

Overall, these results confirm the necessity to develop dynamical models that directly involve behaviours, to gain more precise insight.

\begin{figure}[t]
	\centering
	\includegraphics[width=0.85\linewidth]{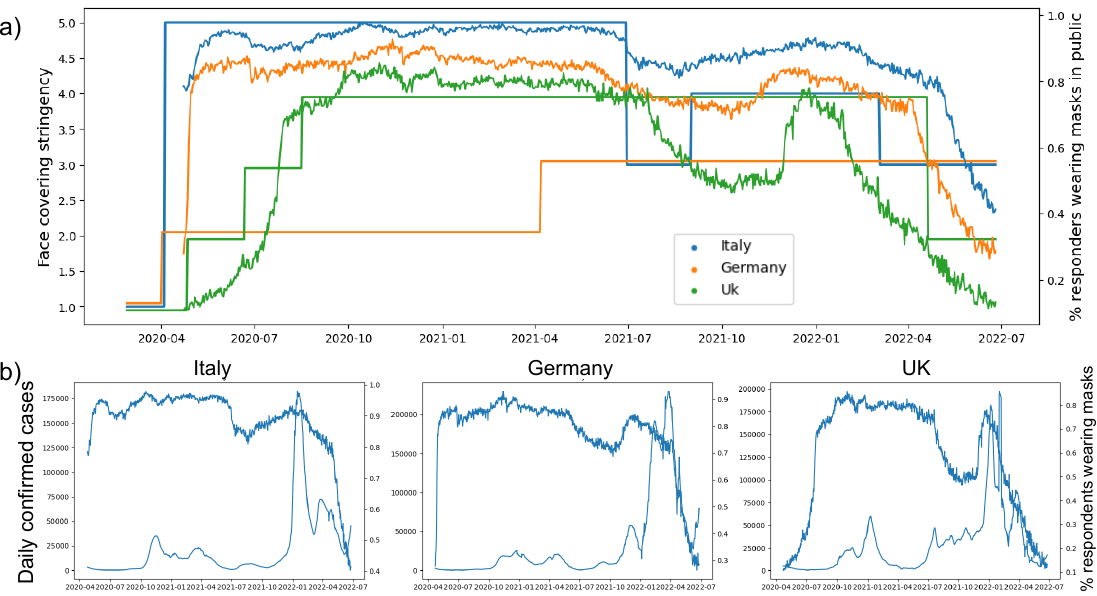}
	\caption{\small Time evolution of mask wearing behaviours and (a) face covering stringency index; (b)  detected case numbers.}
	\label{fig:policy}
\end{figure}

Considering the evolution of behavioural responses with the enforcement of non-pharmaceutical interventions (NPIs) and the disease dynamics, Fig. \ref{fig:policy}a shows the evolution of face covering stringency (from 1 = ``no policy'', to 5 = ``required outside-the-home at all time'') prescribed by the authorities at different times. Overall, the behavioural trends follow closely the evolution of stringency values, in particular during the early pandemic phase. Moreover, Fig. \ref{fig:policy}b shows the time evolution of mask-wearing behaviours and detected case numbers.

Comparing Fig. \ref{fig:policy}a and Fig. \ref{fig:policy}b can inform the development of a dynamical behavioural-epidemic model, and set a range of magnitude for its parameters. We observe the co-evolution of behavioural patterns, proportional to case numbers, at late pandemic stages, after the most stringent measures are lifted, and with cases significantly and rapidly decreasing. In these cases, we may observe pandemic ``fatigue'', i.e., the tendency to stop complying with NPIs after a long period of stringent measures \cite{Montefusco2022}. 
At the population level, fatigue starts being observed after a relatively long time, towards the decreasing phase of the last wave. Instead, previous relaxations of behaviours can be explained by the relaxation of policy measures: we speculate that NPIs and policy measures are the main drivers of behavioural trends, in particular during the early pandemic phases. As a consequence, models should include both mechanisms for mean-field, global interventions, and peer-to-peer mechanisms yielding dynamical effects through mixing. Moreover, most parameters, including those related to behaviours, should be dependent on the pandemic phase. Finally, epidemic and behavioural time scales are clearly commensurable, and thus no separation of time scales can be performed. This requires more complex models, which may give rise to intriguing dynamics. As a final note, we recall that most participants included in the analysis reported not showing COVID-19 symptoms; hence, the curves in the figures above mostly relate to ``susceptible'' individuals.

\section{Behavioural-epidemiological modelling}

Building on the results from data analysis, we develop a dynamical compartment-based behavioural-epidemiological model and analyse its key features for pandemic monitoring.

\subsection{Model development}

Building on \cite{Bulai2023,Peng2021} and the literature on multi-layered compartment models, as well as on the previous data analysis, we construct a dynamical model by intertwining two layers that respectively capture disease spreading and behavioural dynamics. The disease layer has a classic SIRS-model structure, and thus captures the essential features of COVID-19 dynamics \cite{Proverbio2024}. The behavioural layer has three compartments: Heedless, $H$ (who behave without much care about guidelines); Compliant, $C$ (who actively try to avoid becoming infected or infecting); Against, $A$ (who do not follow risk-mitigating guidelines and do not adopt aligned behaviours as the pandemic unfolds). In fact, from the data analysis, it is clear that people who are compliant ($C$) with mask wearing recommendations are never 100\%; hence the introduction of compartments for non-compliant behaviours, either Heedless ($H$) or explicitly Against ($A$). 

Behaviours characterise all stages of the SIRS model. By looking at data and literature \cite{Usher2020}, we assume that Headless behaviour characterises the early stages of the pandemic, but that people eventually lean towards Compliant or Against behaviours. Still, changes in behaviour may also occur during the epidemic progression, either due to peer-pressure (depending on the size of the opposite group and on its level of ``persuasion'') or due to ``fatigue''. We also include the possibility of waning immunity \cite{Kemp2021}, with newly susceptible people re-distributing into each behaviour with a certain probability $\phi$.

Coupling behaviour structure and SIRS epidemic model results in seven mutually exclusive dynamical compartments that satisfy conservation of the total population. The variables representing population fractions in each compartment dynamically evolve according to the ordinary-differential-equation system
\begin{footnotesize}
    \begin{equation}
    \begin{cases}
        \dot{S}_H &= - \psi k_1 S_H  C - k_2 S_H  A + \lambda_1 S_C + \lambda_2 S_A + \delta (1 - \phi) R_C - \beta S_H  I \\
        \dot{S}_C &= \psi k_1 S_H  C + \delta \phi R_C - \lambda_1 S_C - \beta \rho S_C  I \\
        \dot{S}_A &= k_2 S_H  A - \lambda_2 S_A - \beta S_A I + \delta R_A \\
        \dot{I}_C &= \beta \rho S_C  I + \beta S_H  I + \psi k_3 I_A  C - \lambda_3 I_C - k_4 I_C  A + \lambda_4 I_A - \gamma I_C \\
        \dot{I}_A &= \beta S_A  I - \psi k_3 I_A  C + \lambda_3 I_C + k_4 I_C  A - \lambda_4 I_A - \gamma I_A \\
        \dot{R}_C &= \gamma I_C - k_6 R_C  A + \lambda_6 R_A + \psi k_5 R_A  C - \lambda_5 R_C - \delta R_C \\
        \dot{R}_A &= \gamma I_A + k_6 R_C  A - \lambda_6 R_A - \psi k_5 R_A  C + \lambda_5 R_C - \delta R_A
    \end{cases}
    \label{eq:model}
\end{equation}
\end{footnotesize}

where $A = S_A + I_A + R_A$ is the total fraction of Against, $C = S_C + I_C + R_C$ is the total fraction of Compliant, $I = \epsilon I_C + I_A$ is the total fraction of individuals contributing to the infection process ($\epsilon < 1$ captures the fraction of compliant infectious individuals who contribute to the spreading dynamics). Parameters $k_i$, $i=1 \dots 6$, represent the intensity of the peer pressure to change behaviour; $\lambda_i$ capture ``fatigue'', i.e., the rate at which behaviour is changed due to the inability to sustain it for prolonged times. Our data analysis suggests $\lambda_i \ll k_i, \forall i$. Parameter $\psi$ represents an increased incentive for individuals to join Compliant compartments, and can be regarded as a first mean-field intervention at a government level; $\rho$ is a protection factor reducing the risk of Compliant people to become infected; $\beta$ is the disease infection rate; $\gamma$ is the recovery rate; $\delta$ is the rate at which immunity wanes, over longer time scales than infection dynamics. Parameter values may change in different epidemic phases.

\subsection{Computing the basic reproduction number}

\begin{figure}[t]
	\centering
	\includegraphics[width=0.9\linewidth]{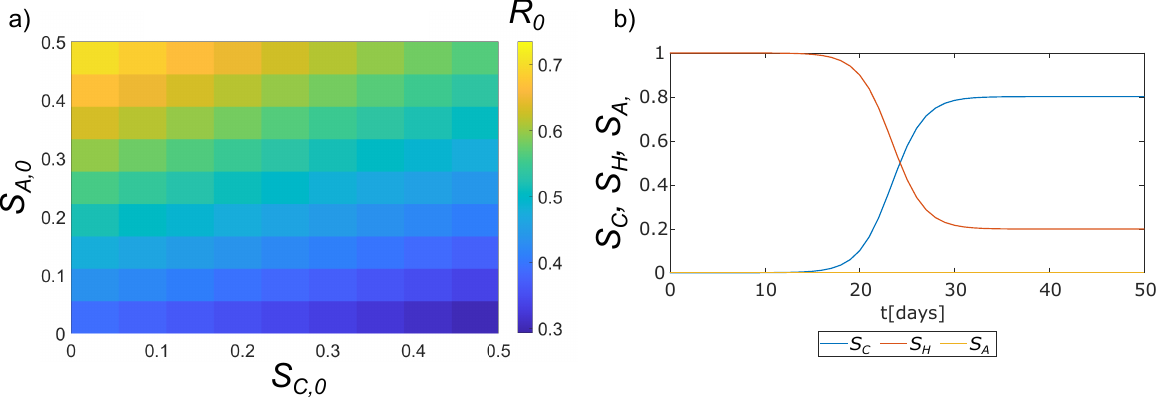}
	\caption{\small (a) Value of $R_0$ depending on the initial conditions $S_{C,0}$ and $S_{A,0}$. (b) Dynamics of the susceptible compartments, when $R_0 = 0.4$.}
	\label{fig:R0}
\end{figure}

At the onset of an epidemic, its rate of unfolding is quantified by the basic reproduction number $R_0$, which acts as a bifurcation parameter governing whether the epidemic bursts ($R_0 > 1$), vanishes ($R_0 < 1$) or remains endemic ($R_0~=~1$) \cite{Kemp2021}. For generalised SIR models, $R_0$ is defined as the spectral radius of the next-generation matrix \cite{Driessche2017}. Consider a multi-compartment generalised SIR epidemiological model (akin to those in \cite{Arino2007,CKG2024}) with state vector $x = \begin{bmatrix}x_S^\top & x_I^\top & x_R^\top \end{bmatrix}^\top$, where vector $x_S \in \mathbb{R}^l$ stacks the fractions of population in susceptible compartments, vector $x_I \in \mathbb{R}^m$ the fractions of population in infectious compartments, and vector $x_R \in \mathbb{R}^n$ the fractions of population in removed compartments.
Assume that the disease-free equilibrium $x_0 = \begin{bmatrix}\bar x_S^\top & 0 & \bar x_R^\top \end{bmatrix}^\top$ (with $\bar x_I \equiv 0$) exists and is stable in the absence of contagion, and that the linearised equations for variables $x_I$ at the disease-free equilibrium are decoupled from all the other system equations.
Then, following \cite{Driessche2017}, the equations for the infectious variables $x_I$ can be rewritten as
\begin{equation}\label{eq:compact}
\dot{x}_{I,i} = \mathcal{F}_i(x) - \mathcal{V}_i(x), \quad i \in \{1,\dots,m\},
\end{equation}
where $\mathcal{F}_i(x)$ is the rate of appearance of new infections in the infected compartment $i$, while $\mathcal{V}_i(x)$ is the rate of other transitions between the infected compartment $i$ and other infected compartments; see \cite{Driessche2017} for details.
Then, if we define the matrices
\begin{equation}
F = \Big[\frac{\partial\mathcal{F}_i(x)}{\partial x_{I,j}} \Big]_{x=x_0} \quad \mbox{and} \quad V = \Big[\frac{\partial\mathcal{V}_i(x)}{\partial x_{I,j}} \Big]_{x=x_0}, \quad 1 \leq i,j \leq m,
\end{equation}
corresponding to the Jacobian of functions $\mathcal{F}$ and $\mathcal{V}$ evaluated at the disease-free equilibrium, the next generation matrix is $F V^{-1}$ and the basic reproduction number can be computed as \cite{Driessche2017}
\begin{equation}
    R_0 = r(F \, V^{-1}) ,
\end{equation}
where $r(M)$ denotes the spectral radius of matrix $M$. In our case, system \eqref{eq:model} can be written in the form \eqref{eq:compact} with $x_I = \begin{bmatrix}
    I_C \\
    I_A 
    \end{bmatrix}$, and then
\begin{equation}
F = \beta \begin{bmatrix}
 \varepsilon(\rho S_C + S_H) & \rho S_C + S_H\\
    \varepsilon S_A & S_A 
    \end{bmatrix}_{x=x_0}
\end{equation}
and
\begin{equation}
V = \begin{bmatrix}
    \lambda_3+k_4(S_A+I_A+R_A)+\gamma-\psi k_3I_A & k_4I_C-\lambda_4-\psi k_3(S_C+I_C+R_C)\\
    \psi k_3I_A - \lambda_3-k_4(S_A+I_A+R_A)  & \psi k_3(S_C+I_C+R_C)-k_4I_C+\lambda_4+\gamma
    \end{bmatrix}_{x=x_0}
\end{equation}
and thus:
\begin{footnotesize}
\begin{equation}
    R_0 =\left. \frac{\beta}{\gamma}\frac{Z}{\lambda_3 + \lambda_4 + \gamma + k_4(S_A+I_A+R_A-I_C) + \psi k_3 (S_C+I_C+R_C-I_A)}\right|_{x=x_0}
    \label{eq:R_0final}
\end{equation}
\end{footnotesize}
where $Z=S_A(\gamma + \lambda_3+\varepsilon \lambda_4) + (S_H + \rho S_C)(\lambda_3+\varepsilon \gamma + \varepsilon\lambda_4) + (S_A+S_H+\rho S_C)[(I_A - \varepsilon I_C)(k_4- \psi k_3)+k_4(R_A+S_A)+ \psi\varepsilon k_3(R_C+S_C)]$.
We can immediately understand how the fractions of compliant, heedless and against population at the initial time affect the pandemic evolution, by looking at their contribution to the basic reproduction number: their proportion may be such that $R_0<1$ and thus suppress the epidemic. For the simple disease-free equilibrium with $\bar S_H=1$, we have $R_0 = \frac{\beta(\lambda_3+\varepsilon(\lambda_4 + \gamma))}{\gamma (\lambda_3+\lambda_4 + \gamma)}$. An example is provided by Fig. \ref{fig:R0}a, where $\beta = 0.4$, $\gamma = 0.35$, $\rho = 0.65$, $\psi=1$, $\epsilon = 0.15$, $k_3 = 0.5$, $k_4 = 0.243$, $\lambda_3 = 0.143$, $\lambda_4 = 0.143$ are chosen as realistic values. The initial conditions for infected compartments are set almost at zero and for recovered compartments at zero, to mimic realistic scenarios at the beginning of an epidemic. 

From Eq. \eqref{eq:R_0final}, we also notice the balance between the terms $\psi k_3$ and $k_4$, which capture the tendency to change from one behaviour to the other. Mean-field interventions $\psi$ tune the individuals' innate propensity to move towards compliant behaviours. An example of the effect of terms $\psi k_3$ and $k_4$ can be seen in the time evolution of the Susceptible compartments, shown in Fig. \ref{fig:R0}b, starting from the initial conditions $S_{C,0} = 10^{-6}$, $S_{A,0} = 10^{-6}$ and $S_{H,0} = 1- S_{A,0} -S_{C,0}$. The asymmetry between persuasion parameters turns most people into being compliant and the epidemic dies out because $R_0=0.4 < 1$.

\section{Discussion and Conclusion}

This work highlighted the need to explicitly account for behaviours when integrating social aspects into epidemiological modelling, in order to better understand epidemic dynamics. The data analysis elucidates the relationship between behaviours and other personal aspects like awareness and trust, as well as the connections between guidelines and peer pressure.
The analysis is univariate, to investigate and develop hypotheses about single aspects related to model development. Future work may consider employing multivariate analysis to corroborate the present findings and extend the scope of the results. Overall, the proposed model integrates various data-informed aspects, related to variable relationships and parameter ranges. It allows us to obtain quantitative insight into the role of compliant behaviours, and paves the way for further analysis of stability properties, sensitivity to control interventions, and adherence to empirical data. Building on our preliminary results, future work will extend the model results to test hypotheses and predictions.

\subsection*{Acknowledgments}
The authors would like to thank dr. M.T.J. Heino for his valuable insights about awareness and behavioural indicators. This work was funded by the European Union through the ERC INSPIRE grant (project number 101076926). Views and opinions expressed are however those of the authors only and do not necessarily reflect those of the European Union or the European Research Council. Neither the European Union nor the granting authority can be held responsible for them.


%
%


\begin{thebibliography}{6}


\bibitem{Alamo2021}
Alamo, T., Reina, D. G., Gata, P. M., Preciado, V. M., Giordano, G.: Data-driven methods for present and future pandemics: Monitoring, modelling and managing. Annu. Rev. Contr., 52, 448-464 (2021). 


\bibitem{Anderson2019}
Anderson, B.D.O., Ye, M.: Recent Advances in the Modelling and Analysis of Opinion dynamics on Influence Networks. Int. J. Autom. Comput. 16(2), 129–149 (2019). 

\bibitem{Arino2007}
Arino, J., Brauer, F., van den Driessche, P., Watmough, J., Wu, J. A final size relation for epidemic models. Math. Biosci. Eng., 4(2), 159-175 (2007). 

\bibitem{Astley2021}
Astley, C. M., Tuli, G., Mc Cord, K. A., Cohn, E. L., Rader, B., Varrelman, T. J., and others: Global monitoring of the impact of the COVID-19 pandemic through online surveys sampled from the Facebook user base. Proc. Natl. Acad. Sci. USA, 118(51), e2111455118 (2021). 

\bibitem{Barkay2020}
Barkay, N., Cobb, C., Eilat, R., Galili, T., Haimovich, D., LaRocca, S., and others: Weights and methodology brief for the COVID-19 symptom survey by University of Maryland and Carnegie Mellon University, in partnership with Facebook. arXiv (2020). 


\bibitem{Bauch2013}
Bauch, C., d’Onofrio, A., Manfredi, P.: Behavioral epidemiology of infectious diseases: an overview. In ``Modeling the interplay between human behavior and the spread of infectious diseases", 1-19 (2013). ISBN: 978-1-4614-5473-1

\bibitem{Bedson2021}
Bedson, J., Skrip, L. A., Pedi, D., Abramowitz, S., Carter, S., Jalloh, M. F., and others: A review and agenda for integrated disease models including social and behavioural factors. Nat. Hum Behav., 5(7), 834-846 (2021). 

\bibitem{Brauer2017}
Brauer, F.: Mathematical epidemiology: Past, present, and future. Infect. Dis. Mod., 2(2), 113-127 (2017). 

\bibitem{Bulai2023}
Bulai, I.M., Sensi, M., Sottile, S.: A geometric analysis of the SIRS compartmental model with fast information and misinformation spreading. Chaos, Solitons \& Fractals 185: 115104 (2024).

\bibitem{Burzyński2021}
Burzyński, M., Machado, J., Aalto, A., Beine, M., Goncalves, J., Haas, T., and others: COVID-19 crisis management in Luxembourg: Insights from an epidemionomic approach. Econom. Human Biol., 43, 101051 (2021). 

\bibitem{CKG2024}
Calà Campana, F., Katz, R., Giordano, G.: Sequential-Quadratic-Hamiltonian optimal control of epidemic models
with an arbitrary number of infected and non-infected compartments. IEEE Control Systems Letters (2024).


\bibitem{Driessche2017}
van den Driessche, P. Reproduction numbers of infectious disease models. In: Infectious Disease Modelling 2.3, pp. 288–303 (2017)

\bibitem{dong2020interactive}
Dong, E., Du, H., Gardner, L.: An interactive web-based dashboard to track COVID-19 in real time. Lancet Inf. Dis., 20 (5), 533--534 (2020). 

\bibitem{Fenichel2011}
Fenichel, E.P., Castillo-Chavez, C., Ceddia, M.G., Chowell, G., Parra, P.A.G., Hickling, G.J., and others: Adaptive human behavior in epidemiological models. Proc. Natl. Acad. Sci. USA, 108(15),
6306–6311 (2011). 

\bibitem{Frieswijk2022}
Frieswijk, K., Zino, L., Ye, M., Rizzo, A., Cao, M.: A mean-field analysis of a network behavioral–epidemic model. IEEE Cont. Sys. Lett., 6, 2533-2538 (2022). 


\bibitem{Giordano2020}
Giordano, G., Blanchini, F., Bruno, R., Colaneri, P., Di Filippo, A., Di Matteo, A., Colaneri, M.: Modelling the COVID-19 epidemic and implementation of population-wide interventions in Italy. Nat. Med., 26(6), 855-860 (2020). 

\bibitem{Gosak2021}
Gosak, M., Kraemer, M. U., Nax, H. H., Perc, M., Pradelski, B. S.: Endogenous social distancing and its underappreciated impact on the epidemic curve. Sci. Rep., 11(1), 3093 (2021). 

\bibitem{Hale2021}
Hale, T., Angrist, N., Goldszmidt, R., Kira, B., Petherick, A., Phillips, T., and others: A global panel database of pandemic policies (Oxford COVID-19 Government Response Tracker). Nature Human Behav, 5(4), 529-538 (2021). 

\bibitem{Heino2023}
Heino, M.T.J., Proverbio, D., Marchand, G., Resnicow, K., Hankonen, N.: Attractor landscapes: a unifying conceptual model for understanding behaviour change across scales of observation. Health Psy. Rev., 17(4), 655-672 (2023). 

\bibitem{Vargas2022}
Hernandez-Vargas, E. A., González, A. H., Beck, C. L., Bi, X., Campana, F. C., Giordano, G.: Modelling and Control of Epidemics Across Scales. 61st Conf. Dec. Contr. (CDC), 4963-4980 (2022). 

\bibitem{Huys2011}
Huys, R., Jirsa, V.K. (eds.): Nonlinear Dynamics in Human Behavior. Springer Berlin Heidelberg (2011). 

\bibitem{Kemp2021}
Kemp, F., Proverbio, D., Aalto, A., Mombaerts, L., d’Hérouël, A. F., Husch, A., and others: Modelling COVID-19 dynamics and potential for herd immunity by vaccination in Austria, Luxembourg and Sweden. J. Theo. Biol., 530, 110874 (2021). 

\bibitem{Kermak1927}
Kermack, W.O., McKendrick, A.G.: A contribution to the mathematical theory of epidemics. Proc. Roy Soc. A, 115(772):700–721 (1927). 


\bibitem{Montefusco2022}
Montefusco, F., Procopio, A., Bulai, I. M., Amato, F., Pedersen, M. G., Cosentino, C. (2022). Interacting with COVID-19: How population behavior, feedback and memory shaped recurrent waves of the epidemic. IEEE Control Systems Letters, 7, 583-588.

\bibitem{Nunner2021}
Nunner, H., Buskens, V., Kretzschmar, M. A model for the co-evolution of dynamic social networks and infectious disease dynamics. Comput. Soc. Netw., 8(1), 19 (2021). 

\bibitem{Palomo2022}
Palomo-Briones, G. A., Siller, M., Grignard, A.: An agent-based model of the dual causality between individual and collective behaviors in an epidemic. Comput. Biol. Med., 141, 104995 (2022). 

\bibitem{Peng2021}
Peng, K., Lu, Z., Lin, V., Lindstrom, M. R., Parkinson, C., Wang, C., and others. A multilayer network model of the coevolution of the spread of a disease and competing opinions. Math. Mod. Meth. Appl. Sci., 31(12), 2455-2494 (2021). 


\bibitem{Proverbio2021}
Proverbio, D., Kemp, F., Magni, S., Husch, A., Aalto, A., Mombaerts, L., and others. Dynamical SPQEIR model assesses the effectiveness of non-pharmaceutical interventions against COVID-19 epidemic outbreaks. PloS ONE, 16(5), e0252019 (2021). 

\bibitem{Proverbio2024}
Proverbio, D, Kemp, F, Gonçalves, J.: Early warning of SARS-CoV-2 infection. In ``Features, Transmission, Detection, and Case Studies in COVID-19", Elsevier (2024). ISBN: 9780323956475

\bibitem{Saha2020}
Saha, S., Samanta, G. P., Nieto, J. J.: Epidemic model of COVID-19 outbreak by inducing behavioural response in population. Nonlin. Dyn., 102, 455-487 (2020). 

\bibitem{Tyson2020}
Tyson, R. C., Hamilton, S. D., Lo, A. S., Baumgaertner, B. O., Krone, S. M.: The timing and nature of behavioural responses affect the course of an epidemic. Bul. Math. Biol., 82, 1-28 (2020). 

\bibitem{Usher2020}
Usher, K., Jackson, D., Durkin, J., Gyamfi, N., Bhullar, N.: Pandemic‐related behaviours and psychological outcomes; A rapid literature review to explain COVID‐19 behaviours. Int. J. Mental Health Nurs., 29(6), 1018-1034 (2020). 



\bibitem{Ye2021}
Ye, M., Zino, L., Rizzo, A., Cao, M.: Game-theoretic modeling of collective decision making
during epidemics. Phys. Rev. E 104(2), 024314 (2021). 

\bibitem{Zino2021}
Zino, L., Cao, M.: Analysis, prediction, and control of epidemics: A survey from scalar to dynamic network models. IEEE Circuits Syst. Mag. 21(4), 4–23 (2021). 

\end{thebibliography}
\end{document}